\documentstyle[11pt]{article}
\newcommand{\noi}{\noindent}
\newcommand{\be}{\begin{equation}}
\newcommand{\ee}{\end{equation}}
\newcommand{\bea}{\begin{eqnarray}}
\newcommand{\eea}{\end{eqnarray}}
\newcommand{\ba}{\begin{array}}
\newcommand{\ea}{\end{array}}

\title{\large\bf      COSMIC STRINGS FROM\\
		      N= 2, D= 5 SUPERGRAVITY}

\author {\normalsize\bf
    Luis O. Pimentel\thanks{E-mail: lopr@xanum.uam.mx}\\
\normalsize\it Universidad Aut\'onoma Metropolitana-Iztapalapa, \\
\normalsize\it Av. Michoac\'an y Pur\'{\i}sima, Col. Vicentina, Apdo.
	       Postal 55-534, \\
\normalsize\it M\'exico D.F., CP 09340, MEXICO.}

\date{}

\begin{document}

\maketitle

\abstract
Exact solutions of N=2 supergravity in five dimensions are found in the
metric with cylindrical symmetry, a particular case corresponds to  the
exterior of a cosmic string.

To appear in Modern Physics Letters A
\newpage
\section{Introduction}

In a recent work ${}^1$ the possibility of obtaining the
cylindrically solutions to four-dimensional low-energy limit of
string theory was investigated. In this work I consider the same
problem for N=2, D=5 supergravity theory.
During the last years there has been intensive studies of physical theories
in more than four dimensions, among them gravity with torsion ${}^2$,strings,
 superstrings and
supergravity. Recently Balbinot et al. ${}^{3,6}$ have considered cosmological
solutions for the N=2 and D=5 supergravity theory ${}^{4, 5}$ which contains
the
metric $g_{M N}$, a spin$-{3\over 2}$ field $\psi^a_M$ (a =1, 2 is an
internal
index) and a 1-form $B_M$ (M, N = 1, ..., 5 and $\mu \nu$ =1,..., 4). Looking
for a "ground-state" configuration they have set the fermion field and the
electromagnetic field in the metric $g_{M N} $ equal to zero. They have
assumed a local structure $V_4  \times S^1 $, where $V_4$ is a four
dimensional spacetime and they found exact solutions when $V_4$ is the flat
Robertson-Walker spacetime. In a previous paper I have considered the
Bianchi type-I spacetime. Here I
want to consider this theory in the exterior of a cosmic string.
\bigskip
\section{ Field Equations}
\bigskip
The Theory studied by Balbinot et al. ${}^3$ has the five-dimensional line
element

\be
 d S^2 = g_{\mu\nu}(x^{\mu})dx^\mu dx^\nu -\phi ^2 (x^\mu )(d x^5)^2
=d s^2-\phi ^2 (x^\mu )(d x^5)^2.
\label{eq:metr1}
\ee

It is also assumed that the 1-form $B_M$ is $B_M=(0,0,0,0,\psi (x^\mu )\quad)$
. With all these assumptions the theory is equivalent to one with a four
dimensional lagrangian given by

\be
{\cal L} = {\frac{1}{4}}\sqrt{g} \phi \left(R+2{ D_\lambda \psi D^\lambda \psi
\over \phi^2}\right).
\label{eq:lag}
\ee

The field equations obtained from the variation of the above Lagrangian are

\be
G_{\mu \nu}=R_{\mu \nu}-{\frac{1}{2}}g_{\mu\nu}R
={3\over 2 \phi^{2}}(\psi_{,\mu}\psi_{,\nu}
-{\frac{1}{2}}g_{\mu\nu}\psi_{,\lambda}\psi^{,\lambda})+{1\over\phi}
(\phi_{;\mu\nu}-
g_{\mu\nu}\Box \, \phi).
\label{feq1}
\ee

\be
\Box \,\phi +{ \psi _ \lambda \psi ^\lambda \over \phi}
=0,
\label{eq:eqfi}
\ee

\be
\Box \,\psi -{ \phi _ \lambda \psi ^\lambda \over \phi}
=0.
\label{eq:eqpsi}
\ee

In this section we set the field equations for a metric that corresponds
to the exterior of a cosmic string. Here we are interested in a
static infinite-lenght cosmic string, the fields must have static
cylindrical symmetry. The spacetime must have three commuting
Killing vectors fields, one of them timelike and the other two
spacelike, one of them with closed orbits, such that any two
are orthogonal to each other and each is hypersurface orthogonal.
There is an axis where the Killing vector with closed orbits
vanishes. The normalization is chosen so that along any closed
integral curve the parameter takes the values from 0 to 2$\pi$,
 and the norm of the spacelike and timelike Killing vectors is 1
and -1, respectively, on the axis. Using coordinates t,z,$\rho$
and $\theta$, where $\rho$ is the geodesic distance from the axis
in the direction orthogonal to the three Killing vector fields,
$(\partial /\partial t)^{a}$ is the timelike Killing field , and
$(\partial /\partial \theta)^{a} $is the Killing
 field that has closed orbits, and $(\partial /\partial z)^{a} $ is
the Killing vector field
along the axis. With all these assumptions the metric is

\be
ds^2=-e^{2A}dt^2+e^{2B}dz^2+ e^{2C}d\theta ^2+d\rho ^2,
\label{eq:metr6}
\ee
\noi where A, B, and C are functions of the radial distance $\rho$.
The range of $\theta$ is $[0,2\pi]$ and $\theta =0$ and
$\theta=2\pi$ are identified. The boundary conditions on the axis
of the cosmic strigs are

\bea
\lim_{\rho \rightarrow 0} A(\rho)&= \lim_{\rho \rightarrow 0} B(\rho)=0,\\
\nonumber\\
\lim_{\rho \rightarrow 0} e^{2C}/\rho^2 &=1.
\label{eq:bound}
\eea

In order
to make the algebra simpler we will use here the following metric,
\be
ds^2=-e^{2A}dt^2+e^{2B}dz^2+ e^{2C}d\theta ^2+e^{2(A+B+C)}dr ^2,
\label{eq:metr}
\ee
\noi where A, B, and C are now functions of the new radial distance r,
and the transformation is given by
\be
d\rho =e^{A+B+C} dr.
\label{eq:transf}
\ee
The
field equations for this particular spacetime are given by

\be
B'' + C'' -A'B'-A'C'-B'C' +\frac{\phi ''}{\phi}-A'\frac{\phi
'}{\phi}+\frac{3}{4}\left
(\frac{\psi '}{\phi} \right )^2 =0
\label{eq:q1}
\ee

\be
(A'+B'+C')\frac{\phi
'}{\phi}+A'B'+A'C'+B'C'- \frac{3}{4}\left
(\frac{\psi '}{\phi} \right )^2 =0
\label{eq:q2}
\ee

\be
A''+ C'' -A'B'-A'C'-B'C' +\frac{\phi ''}{\phi}-B'\frac{\phi
'}{\phi}+\frac{3}{4}\left
(\frac{\psi '}{\phi} \right )^2 =0
\label{eq:q3}
\ee

\be
A''+B''  -A'B'-A'C'-B'C' +\frac{\phi ''}{\phi}-C'\frac{\phi
'}{\phi}+\frac{3}{4}\left
(\frac{\psi '}{\phi} \right )^2 =0
\label{eq:q4}
\ee

\noindent and
\be
 \phi ''+\frac{ {\psi'}^2 }{\phi}
=0,
\label{eq:q5}
\ee

\be
 \psi '' -\frac{{ \psi ' \phi'} }{\phi}
=0,
\label{eq:q6}
\ee

\noi here the prime derivative with respect to r. In the following
section we look for exact solutions to the system of
differential equations (7 -12).
\bigskip
\section{\bf  Exact Solutions}
\bigskip
{}From Eqs.(\ref{eq:q5} and \ref{eq:q6}) it follows that
\be
\phi =L \cos(\omega r+\delta_0 ),
\label{eq:solphi}
\ee
\be
\psi= \omega \int{\phi}=L\sin(\omega r+\delta_0).
\label{eq:solpsi}
\ee
\noi where L, $\omega$, and $\delta_0$ are integration constants.
Adding Eqs.(\ref{eq:q2}and \ref{eq:q4}) we have
\be
{{\phi''}\over{\phi}}+{{\phi'}\over{\phi}}(A+B)'+(A+B)''=0
\ee
\noi and the solution is
\be
A+B=q_1\int{\frac{dr'}{\phi(r')}}-\log(\phi),
\ee
\label{eq:ab}
\noi where $q_1$ is an integration constant and the other
integration constant was set to zero without loss of generality. Adding
Eqs.(\ref{eq:q2} and\ref{eq:q3}) and Eqs.(\ref{eq:q1} and
\ref{eq:q2}) we obtain identical differential equations
for A+C and B+C with the corresponding solutions,
\be
A+C=q_2\int{\frac{dr'}{\phi(r')}}-\log(\phi),
\ee
\label{eq:ac}
\be
B+C=q_3\int{\frac{dr'}{\phi(r')}}-\log(\phi),
\ee
\label{eq:bc}
{}From the above expressions we obtain the explicit solutions for A, B and C,

\be
A= a_1\log[\tan(\omega r+\delta_0) +\sec(\omega r+\delta))]-\frac{1}{2}
\log [\cos(\omega r+\delta_0)]
\ee
\label{eq:sola}


\be
B= b_1\log[\tan(\omega r+\delta_0) +\sec(\omega r+\delta)]-\frac{1}{2}
\log [\cos(\omega r+\delta_0)]
\ee
\label{eq:solb}

\be
C= c_1\log[\tan(\omega r+\delta_0) +\sec(\omega r+\delta_0)]-\frac{1}{2}
\log [\cos(\omega r+\delta_0)]
\ee
\label{eq:solc}

\noi where $a_1, b_1$, and $c_1$ are integration constants that
satisfy the following relation

\be
a_1b_1+a_1c_1+b_1c_1=\frac{3}{4}.
\ee
\label{eq:}
The local 'deficit angle' is of the form

\be
\delta (r)= 2\pi \bigl ( 1-\frac{1}{\sqrt{g_{r\, r}}}\frac{\partial}{\partial
r}\sqrt{g_{\theta \theta}} \bigr
)= 2 \pi \Bigl [1-\frac{\omega c_1 \sec(\omega r+\delta_0)}{
\bigl (\tan(\omega r+\delta_0)+\sec(\omega r+\delta_0)\bigr )^{(a_1
+b_1)}}\Bigr ] ,
\ee

\noi that is non-vanishing in general. The case of the cosmic string
corresponds to $A(r)=B(r)$, or $a_1=b_1$; that is, the cosmic
string exhibits explicit Lorentz invariance along its axis.

\section{ Acknowledgment}
This work was partially supported by CONACYT grant 1861-E9212.



\begin{thebibliography}{99}

\bibitem{1.} M. J. Rogatko, {\sl Phys. A: Math. Gen.} {\bf 26}, L575 (1993).

\bibitem{2.} G. German, A. Macias and O. Obregon, {\sl Class. Quantum
Grav.},{\bf 10},  1045 (1993).
\bibitem{3.}  R. Balbinot, J.C. Fabris and R. Kerner, {\sl Class. Quantum.
Grav.}
	  {\bf 7}, L17 (1990).

\bibitem{4.} R. D'Auria, P. Fre, E. Maine and T. Regge, {\sl Ann. Phys. NY }
	 {\bf{135}}, 237 (1982).

\bibitem{5.} E. Cremmer, {\sl Superspace and Supergravity, Proceedings of the
	 Nuffield Workshop (Cambridge: University Press)} 1980.

\bibitem{6.} R. Balbinot, J. C. Fabris, and R. Kerner, {\sl Phys. Rev. D} {\bf
42}, 1023 (1990).



\end{thebibliography}
\end{document}